\input harvmac
\input epsf
\noblackbox

\newcount\figno
\figno=0
\def\fig#1#2#3{
\par\begingroup\parindent=0pt\leftskip=1cm\rightskip=1cm\parindent=0pt
\baselineskip=11pt
\global\advance\figno by 1
\midinsert
\epsfxsize=#3
\centerline{\epsfbox{#2}}
\vskip 12pt
\centerline{{\bf Figure \the\figno} #1}\par
\endinsert\endgroup\par}
\def\figlabel#1{\xdef#1{\the\figno}}

\def\pano{\par\noindent}

\def\pmb#1{\setbox0=\hbox{#1}%
 \kern-.025em\copy0\kern-\wd0
 \kern.05em\copy0\kern-\wd0
 \kern-.025em\raise.0433em\box0 }
\font\cmss=cmss10
\font\cmsss=cmss10 at 7pt
\def\rlx{\relax\leavevmode}
\def\Cop{\relax\,\hbox{$\inbar\kern-.3em{\rm C}$}}
\def\Rop{\relax{\rm I\kern-.18em R}}
\def\Nop{\relax{\rm I\kern-.18em N}}
\def\Pop{\relax{\rm I\kern-.18em P}}
\def\Zop{\rlx\leavevmode\ifmmode\mathchoice{\hbox{\cmss Z\kern-.4em Z}}
 {\hbox{\cmss Z\kern-.4em Z}}{\lower.9pt\hbox{\cmsss Z\kern-.36em Z}}
 {\lower1.2pt\hbox{\cmsss Z\kern-.36em Z}}\else{\cmss Z\kern-.4em
 Z}\fi}


\def\AdSs5{$AdS_5$}
\def\AdS5s5{$AdS_5 \times S^5$}


\def\eg{{\it e.g.}}

\def\IZ{\relax\ifmmode\mathchoice {\hbox{\cmss Z\kern-.4em
Z}}{\hbox{\cmss Z\kern-.4em Z}} {\lower.9pt\hbox{\cmsss Z\kern-.4em
Z}} {\lower1.2pt\hbox{\cmsss Z\kern-.4em Z}}\else{\cmss Z\kern-.4em
Z}\fi}

\parindent 25pt
\overfullrule=0pt
\tolerance=10000
\def\ie{{\it i.e.}}


\def\pmb#1{\setbox0=\hbox{#1}%
 \kern-.025em\copy0\kern-\wd0
 \kern.05em\copy0\kern-\wd0
 \kern-.025em\raise.0433em\box0 }
\font\cmss=cmss10
\font\cmsss=cmss10 at 7pt
\def\rlx{\relax\leavevmode}
\def\Zop{\rlx\leavevmode\ifmmode\mathchoice{\hbox{\cmss Z\kern-.4em Z}}
 {\hbox{\cmss Z\kern-.4em Z}}{\lower.9pt\hbox{\cmsss Z\kern-.36em Z}}
 {\lower1.2pt\hbox{\cmsss Z\kern-.36em Z}}\else{\cmss Z\kern-.4em
 Z}\fi}


\def\pmb#1{\setbox0=\hbox{#1}%
 \kern-.025em\copy0\kern-\wd0
 \kern.05em\copy0\kern-\wd0
 \kern-.025em\raise.0433em\box0 }
\font\cmss=cmss10
\font\cmsss=cmss10 at 7pt
\def\rlx{\relax\leavevmode}
\def\Cop{\relax\,\hbox{$\kern-.3em{\rm C}$}}
\def\Rop{\relax{\rm I\kern-.18em R}}
\def\Nop{\relax{\rm I\kern-.18em N}}
\def\Pop{\relax{\rm I\kern-.18em P}}

\def\Zop{\rlx\leavevmode\ifmmode\mathchoice{\hbox{\cmss Z\kern-.4em Z}}
 {\hbox{\cmss Z\kern-.4em Z}}{\lower.9pt\hbox{\cmsss Z\kern-.36em Z}}
 {\lower1.2pt\hbox{\cmsss Z\kern-.36em Z}}\else{\cmss Z\kern-.4em
 Z}\fi}

\def\Res{{\rm Res}}
\def\Nbar{\bar{N}}
\def\fbar{\bar{f}}
\def\fhat{\hat{f}}
\def\fcheck{\check{f}}
\def\cf{{\it cf.}}


\lref\metsaev{
R.~R.~Metsaev,
{\it Type IIB Green-Schwarz superstring in plane wave Ramond-Ramond
background},
Nucl.\ Phys.\ B {\bf 625}, 70 (2002); {\tt hep-th/0112044}.
}

\lref\metsaevtseytlin{
R.~R.~Metsaev, A.~A.~Tseytlin,
{\it Exactly solvable model of superstring in plane wave Ramond-Ramond  background,}
Phys.\ Rev.\ D {\bf 65}, 126004 (2002);
{\tt hep-th/0202109}.}

\lref\bmn{
D.~Berenstein, J.~M.~Maldacena, H.~Nastase,
{\it Strings in flat space and pp waves from N = 4 super Yang Mills,}
JHEP {\bf 0204}, 013 (2002);
{\tt hep-th/0202021}.}

\lref\GSstringfield{
M.~B.~Green, J.~H.~Schwarz,
{\it Superstring Field Theory},
Nucl.\ Phys.\ B {\bf 243}, 475 (1984).}

\lref\GSstringint{
M.~B.~Green, J.~H.~Schwarz,
{\it Superstring Interactions,}
Nucl.\ Phys.\ B {\bf 218}, 43 (1983).}

\lref\GSB{
M.~B.~Green, J.~H.~Schwarz, L.~Brink,
{\it Superfield Theory Of Type II Superstrings},
Nucl.\ Phys.\ B {\bf 219}, 437 (1983).
}

\lref\lss{
J.~Lucietti, S.~Schafer-Nameki, A.~Sinha, {\it
On the exact open-closed vertex in plane-wave light-cone string field
theory}, to appear in Phys.\ Rev.\ D;
{\tt hep-th/0311231}.
}

\lref\spradlinvolovich{
M.~Spradlin, A.~Volovich,
{\it Superstring interactions in a pp-wave background,}
Phys.\ Rev.\ D {\bf 66}, 086004 (2002);
{\tt hep-th/0204146}.}

\lref\spradlinvolovichII{
M.~Spradlin, A.~Volovich,
{\it Superstring interactions in a pp-wave background. II},
JHEP {\bf 0301}, 036 (2003); {\tt hep-th/0206073}.}

\lref\gomis{
J.~Gomis, S.~Moriyama, J.~w.~Park,
{\it Open + closed string field theory from gauge fields;}
{\tt hep-th/0305264}.}

\lref\bogdanone{
B.~J.~Stefanski,
{\it Open string plane-wave light-cone superstring field theory;}
{\tt hep-th/0304114};  
B.~Chandrasekhar, A.~Kumar,
{\it D-branes in pp-wave light cone string field theory,}
JHEP {\bf 0306}, 001 (2003);
{\tt hep-th/0303223.}}

\lref\bgg{
O.~Bergman, M.~R.~Gaberdiel, M.~B.~Green,
{\it D-brane interactions in type IIB plane-wave background,}
JHEP {\bf 0303}, 002 (2003);
{\tt hep-th/0205183}.
}

\lref\taka{
T.~Takayanagi,
{\it Modular invariance of strings on pp-waves with RR-flux,}
JHEP {\bf 0212}, 022 (2002);
{\tt hep-th/0206010.}
}

\lref\chandra{
B.~Chandrasekhar, A.~Kumar,
{\it D-branes in pp-wave light cone string field theory,}
JHEP {\bf 0306}, 001 (2003);
{\tt hep-th/0303223}.}

\lref\bgmnn{
D.~Berenstein, E.~Gava, J.~M.~Maldacena, K.~S.~Narain, H.~Nastase,
{\it Open strings on plane waves and their Yang-Mills duals};
{\tt hep-th/0203249}.}

\lref\hssv{
Y.~H.~He, J.~H.~Schwarz, M.~Spradlin, A.~Volovich,
{\it Explicit formulas for Neumann coefficients in the plane-wave geometry,}
Phys.\ Rev.\ D {\bf 67}, 086005 (2003);
{\tt hep-th/0211198,v2}.}

\lref\schwarz{
J.~H.~Schwarz,
{\it Comments on superstring interactions in a plane-wave background},
JHEP {\bf 0209}, 058 (2002);
{\tt hep-th/0208179}.
}

\lref\gg{
M.~R.~Gaberdiel, M.~B.~Green,
{\it The D-instanton and other supersymmetric D-branes in IIB plane-wave string theory,}
Annals Phys.\  {\bf 307}, 147 (2003);
{\tt hep-th/0211122}.}

\lref\ggss{
M.~R.~Gaberdiel, M.~B.~Green, S.~Schafer-Nameki, A.~Sinha,
{\it Oblique and curved D-branes in IIB plane-wave string theory,}
JHEP {\bf 0310}, 052 (2003);
{\tt hep-th/0306056}.}

\lref\asnvs{
A.~Sinha, N.~V.~Suryanarayana,
{\it Tadpole analysis of orientifolded plane-waves,}
JHEP {\bf 0211}, 026 (2002);
{\tt hep-th/0209247}.}

\lref\GSWII{
M.~B.~Green, J.~H.~Schwarz, E.~Witten,
{\it Superstring Theory. Vol. 2: Loop Amplitudes, Anomalies And
Phenomenology}, Cambridge University Press (1987).
}

\lref\ksv{
I.~R.~Klebanov, M.~Spradlin, A.~Volovich,
{\it New effects in gauge theory from pp-wave superstrings},
Phys.\ Lett.\ B {\bf 548}, 111 (2002);
{\tt hep-th/0206221}.}

\lref\minxin{
M.~x.~G.~Huang,
{\it Three point functions of N = 4 super Yang Mills from light cone string
field theory in pp-wave},
Phys.\ Lett.\ B {\bf 542}, 255 (2002);
{\tt hep-th/0205311}.}

\lref\chukhoze{
C.~S.~Chu, V.~V.~Khoze,
{\it Correspondence between the 3-point BMN correlators and the 3-string vertex
on the pp-wave},
JHEP {\bf 0304}, 014 (2003);
{\tt hep-th/0301036}.}

\lref\giantgraviton{
V.~Balasubramanian, M.~x.~Huang, T.~S.~Levi, A.~Naqvi,
{\it Open strings from N = 4 super Yang-Mills},
JHEP {\bf 0208}, 037 (2002); {\tt hep-th/0204196}.} 

\lref\pank{
A.~Pankiewicz,
{\it More comments on superstring interactions in the pp-wave background},
JHEP {\bf 0209}, 056 (2002);
{\tt hep-th/0208209}.}

\lref\pankbog{
A.~Pankiewicz, B.~Stefanski,
{\it Pp-wave light-cone superstring field theory},
Nucl.\ Phys.\ B {\bf 657}, 79 (2003);
{\tt hep-th/0210246}.}

\lref\pankrev{
A.~Pankiewicz,
{\it Strings in plane wave backgrounds},
Fortsch.\ Phys.\  {\bf 51}, 1139 (2003);
{\tt hep-th/0307027}.}

\lref\plefka{
J.~C.~Plefka,
{\it Lectures on the plane-wave string / gauge theory duality},
Fortsch.\ Phys.\  {\bf 52}, 264 (2004); 
{\tt hep-th/0307101}.
}

\lref\spradlinreview{
M.~Spradlin, A.~Volovich,
{\it Light-cone string field theory in a plane wave};
{\tt hep-th/0310033}.
}

\lref\sheik{
D.~Sadri, M.~M.~Sheikh-Jabbari,
{\it The plane-wave / super Yang-Mills duality};
{\tt hep-th/0310119}.
}

\lref\russo{
R.~Russo, A.~Tanzini,
{\it The duality between IIB string theory on pp-wave and N = 4 SYM: A status
report};
{\tt hep-th/0401155}.
}

\lref\berensteinnarain{
D.~Berenstein, E.~Gava, J.~M.~Maldacena, K.~S.~Narain, H.~Nastase,
{\it Open strings on plane waves and their Yang-Mills duals},
{\tt hep-th/0203249}.}

\lref\berenstein{
D.~Berenstein,
{\it Shape and holography: Studies of dual operators to giant gravitons},
Nucl.\ Phys.\ B {\bf 675}, 179 (2003);
{\tt hep-th/0306090}.
}

\lref\GR{
I.~S.~Gradshteyn, I.~M.~Ryzhik, {\it Tables of Integrals,
Series, and Products}, Academic Press (2000).}

\lref\WW{
E.~T.~Whittaker, G.~N.~Watson, {\it A Course of Modern
Analysis}, Cambridge University Press (1927).}

\lref\watsonbessel{
G.~N.~Watson, 
{\it A Treatise on the theory of Bessel functions}, Cambridge
University Press (1966).
}


\Title{\vbox{
\hbox{hep-th/0402185}
\hbox{DAMTP-2004-9}
\hbox{DESY-04-024}
}}
{\vbox{\centerline{On the plane-wave cubic vertex}}}
\centerline{James Lucietti$^{\, \flat}$, Sakura Sch\"afer-Nameki$^{\, \sharp}$ and Aninda Sinha$^{\, \flat}$ }
\bigskip
\centerline{\it $^\flat$DAMTP, University of Cambridge}
\centerline{\it Wilberforce Road, Cambridge CB3 OWA, U.K.}
\centerline{\it $^\sharp$II. Institut f\"ur Theoretische Physik,
University of Hamburg}
\centerline{\it Luruper Chaussee 149, 22761 Hamburg, Germany}
\footnote{}{\tt Email: J.Lucietti, S.Schafer-Nameki, A.Sinha@damtp.cam.ac.uk}

\vskip1.4cm
\centerline{\bf Abstract}
\bigskip
\noindent
The exact bosonic Neumann matrices of the cubic vertex in plane-wave
light-cone string field theory are derived using the contour
integration techniques developed in our earlier paper. This simplifies
the original derivation of the vertex. In particular, the
Neumann matrices are written in terms of $\mu$-deformed Gamma-functions, thus casting them into a form that elegantly generalizes the well-known
flat-space solution. The asymptotics of the
$\mu$-deformed Gamma-functions allow one to determine the large-$\mu$
behaviour of
the Neumann matrices including exponential corrections. We provide an
explicit expression for the first exponential correction and make a
conjecture for the subsequent exponential correction terms.

\bigskip

\Date{2/2004}


\newsec{Introduction}

The BMN-correspondence \refs{\bmn} between the plane-wave limit of IIB string
theory on $AdS_5 \times S^5$ \refs{\metsaev, \metsaevtseytlin}
and a certain sector of the $d=4$, ${\cal N}=4$ SYM theory has proven to be a powerful means in
paving the way towards an improved understanding of the AdS/CFT
correspondence.

Despite the impressive advancements in the analysis of both sides
of this correspondence, some key questions still remain to be
answered. One central point that has so far not been addressed to
full satisfaction is the study of interactions and their dual
interpretation. For instance, given the exact cubic vertex, the string
scattering amplitudes can in principle be calculated for all values of $\mu$. For large
$\mu$ this would allow comparison with perturbative gauge theory
calculations and moreover the knowledge of finite $\mu$
corrections could provide very interesting predictions for finite
$\lambda'$ corrections in the gauge theory.

Some important results in this direction have already been obtained.
On the string theory side, progress has been made towards
understanding the interactions in the framework of
light-cone string field theory. This was first developed for flat space in
\refs{\GSstringfield, \GSstringint, \GSB} and then generalized to the
plane-wave in \refs{\spradlinvolovich,
\spradlinvolovichII, \pank, \pankbog,  \hssv, \gomis, \bogdanone, \lss}. For comparisons
to the gauge theory side see \eg\ \refs{\ksv, \minxin,
\chukhoze}\foot{For a  more complete list of
references see \refs{\pankrev, \plefka, \spradlinreview, \sheik, \russo}.}.
However, quantities that have been
computed for all $\mu$ are still rather rare, and we believe comprise only
the results of \refs{\hssv} for the cubic closed interaction vertex as well as
\refs{\lss} for the open-closed vertex. Furthermore, in order to
perform computations of scattering amplitudes based on these results, it would be advantageous
if the interactions are expressed in a concise form reminiscent of the flat-space results.

The main motivation for the present paper is to re-address the analysis
of the plane-wave cubic vertex. The derivation that we shall
provide for the vertex relies on the contour method for summing
certain infinite series, which was
developed in \refs{\lss}. One of the main merits of this approach is
that it allows a derivation very close to
the one known for the flat space vertex in \refs{\GSstringint}.
The solution to the cubic vertex equation, as in \refs{\lss}, will be written in terms of $\mu$-deformed
Gamma-functions. This gives an elegant
generalization of the flat-space solutions of
\refs{\GSstringint}.

There is a slight discrepancy between the explicit
expression of the Neumann matrices in \refs{\hssv}
and ours. It appears that the only
problem in \refs{\hssv} is the final expression for the Neumann
vectors (\ie, their
equation (52))\foot{In fact it is easy to see that there is some error
in (52) of \refs{\hssv}; simply note that $\phi_{m 3}$ in the equation
in question is
divergent.}. We derive the large $\mu$ asymptotics directly from the exact
expression for the Neumann vectors, which agree with those in
\refs{\hssv}. Despite the discrepancy in the exact expression, this is
not surprising since the asymptotics in \refs{\hssv} 
were not developed from their exact expression for the vertex.  We
will elaborate on this in section 3.3.
We also extend the large-$\mu$ asymptotics by explicitly computing the
first exponential corrections and provide a conjecture for the
subsequent exponential correction terms.

The plan of this paper is as follows. In section 2 we present the derivation of the cubic vertex in flat space
using the contour method for summing series of \refs{\lss}. In section 3 we
generalize this to the cubic vertex in the plane-wave.
In section 4 we use the asymptotics of the $\mu$-deformed
Gamma-functions to derive the large-$\mu$ expansions of the
Neumann matrices. We also comment upon the exponential corrections
appearing in the large-$\mu$ asymptotic expansions and explicitly give
the first term, as well as provide a conjecture for the subsequent terms. We conclude in section 5. There are four
appendices, in which various properties and asymptotics of the
$\mu$-deformed Gamma-functions are derived.


\newsec{The cubic vertex in flat-space}

In this section, as a warm-up, we will use the contour method of
\refs{\lss} in order to derive the well-known
flat-space Neumann matrices $\Nbar_{mn}^{rs}$ for the cubic string
vertex, which originally were determined in \refs{\GSstringint}. 
We shall focus on the bosonic part of the vertex, for which the
standard ansatz is
\eqn\roughvertex{
|V\, \rangle = {\cal N} \exp \left ( 
\half \sum_{r,s=1}^3\, \sum_{m,n=1}^\infty
a_{-m}^{(r)} \Nbar_{mn}^{rs} a_{-n}^{(s)}
+ \sum_{r=1}^3\, \sum_{m=1}^\infty
a_{-m}^{(r)} \Nbar_{m}^{r} {\bf P} + K {\bf P}^2
\right) |\,0\, \rangle \,,
}
where $a_{-m}^{(r)}$ are the normalized oscillators of the $r$-th string
and ${\bf P}^i= 2 p_1^+ p_2^i - 2 p_2^+ p_1^i$.
Geometrical continuity conditions and momentum conservation then imply
constraints upon the Neumann matrices. 
It was shown that the Neumann matrices satisfy
\eqn\Nfactor{
\Nbar_{mn}^{rs} = - {mn\alpha_1 \alpha_2\alpha_3 \over n\alpha_r +
m\alpha_s } \Nbar_m^r \Nbar_n^s \,,
}
where
\eqn\Nbarflat{
\Nbar_m^r = - \left(\left(A^{(r)}\right)^t \Gamma^{-1} B
\right)_m \,, \qquad
K= -{1\over 4} B \Gamma^{-1} B \,,
}
and $\alpha_i = 2 p^{(i)+}$ are the momenta of the various strings,
which are chosen such that $\sum_{i=1}^3 \alpha_i=0$, and $\tau_0=
\sum_{r=1}^3 \alpha_r \log |\alpha_r|$. Without loss of
generality we assume $\alpha_1, \alpha_2 >0$ and $\alpha_3<0$.
The matrices $A$ and $B$ are defined in appendix A and correspond
to various Fourier modes. Further 
\eqn\Gammadef{ 
\Gamma =
\sum_{r=1}^3 A^{(r)} \left(A^{(r)}\right)^t \,. 
} 
To solve for the Neumann matrices, it is clear that it is
sufficient to determine $\Gamma^{-1}B$. This is most conveniently done
by solving the two coupled series
\eqn\flatstepwise{\eqalign{
\sum_{n=1}^\infty \sqrt{n} \fbar_n^{(3)} A_{nm}^{(r)} &=
{\alpha_3\over \alpha_r} \sqrt{m}  \fbar_m^{(r)} \cr
\sum_{r=1}^3 \sum_{n=1}^\infty {\sqrt{n}\over \alpha_r} A_{mn}^{(r)}
\fbar_n^{(r)}& = -B_m\,,}}
for $\fbar^{(r)}_n$. These are related to the Neumann
vectors by 
\eqn\Ntof{
\fbar^{(r)}_m ={ \alpha_r\over \sqrt{m}} \Nbar_{m}^r\,.
}
An additional constraint is
\eqn\othercond{
\sum_{n=1}^\infty \sqrt{n} \fbar_n^{(3)} B_n = -2 {\tau_0\over
 \alpha_1\alpha_2} \,,
}
which comes from the knowledge of the explicit form of $K$ in flat-space\foot{Note, that the RHS
of this equation equals $K$, which in flat-space can be determined using
conformal invariance, whereby one maps the known Neumann functions of the
complex plane to the light-cone string diagram, \cf\ \refs{\GSWII}. Due to the absence of
explicit conformal invariance in the light-cone gauge, this is not
possible anymore
in the plane-wave, so that the condition \othercond\ cannot be used to
derive properties of $f^{(3)}$. This will be discussed in the next section.}.
The equations \flatstepwise\ are equivalent to
\eqnn\fone
\eqnn\ftwo 
\eqnn\fthree 
\eqnn\ffour
$$
\eqalignno{
\sum\limits_{m=1}^\infty {(-1)^m\over m} \sin (m \pi \beta )
\fbar^{(3)}_m  &= \pi {\tau_0\over \alpha_3}  &  \fone \cr
\sum_{m=1}^\infty (-1)^{m+n}m \; {\sin(m \pi\beta)\over n^2 -m^2
\beta^2} \; \fbar^{(3)}_m &= -{\pi \over 2 \beta^2} \fbar_n^{(1)} &
\ftwo\cr
\sum\limits_{m=1}^\infty (-1)^m m \; {\sin(m \pi\beta)\over n^2 -m^2
(1+\beta)^2} \; \fbar^{(3)}_m &= {1\over (\beta+1)^2} {\pi\over 2}
\fbar_n^{(2)} &\fthree \cr
{\beta\over \alpha_1 } \sum\limits_{n=1}^\infty  n (-1)^n
{ \fbar^{(1)}_n  \over n^2 -m^2 \beta^2}
 + {1+\beta \over \alpha_2} \sum\limits_{n=1}^\infty &
n
{ \fbar^{(2)}_n \over n^2 -m^2 (1+\beta)^2} \cr
={\pi \over 2 \alpha_3}& (-1)^m {1\over \sin (m\pi \beta)}
\fbar_m^{(3)}- {\alpha_3\over \alpha_1\alpha_2 } {1\over m^2}\,.
&\ffour
}
$$
Here, we defined $\beta=\alpha_1/\alpha_3 <0$.

We shall now apply the contour method in order to derive the
solutions to these equations. Schematically, the contour method provides one with solutions $f(n)$
to certain coupled series $\sum_n f(n) = F$ for given
$F$ (note these can be interpreted as infinite dimensional matrix equations). The main idea is to map the sum to a contour integral in the
plane so that by Cauchy's theorem 
\eqn\contone{
\sum_{n=1}^{\infty} f(n)+ \sum_k Res_{z=z_k} \pi\cot(\pi z)f(z) =
\lim_{R\to \infty} \oint_{C_{R}} {dz\over 2\pi i} \pi\cot(\pi z) f(z)\,,
}
where $C_R$ is the contour given by a circle of radius $R$ centred
on the origin, not intersecting any poles of the integrand (so
in particular $R \neq 1,2,3...$). If the RHS of \contone\ vanishes,
one can compare the sum over residues with $F$, which allows one to 
infer the poles and zeroes of $f(z)$ (note we will assert that
$f(z)=0$ for $z=0,-1,-2...$). For more details and examples on the contour method we refer the reader to \refs{\lss}.

Applied to the present context, \ie, in order to find the solutions for $\fbar^{(r)}_n$, we map
the sum to a contour integral in the complex plane, \ie, consider
in view of \fone 
\eqn\foneint{\eqalign{ 
\oint {dm\over 2\pi i} \; \pi\cot (\pi m)
{1\over \cos (\pi m)} {\sin (m \pi \beta)\over m} \fbar^{(3)} (m) &\cr
= - \oint {dm\over 2\pi i} \; \Gamma (m+1) \Gamma(-m) &  {\sin (m \pi \beta)\over m}
\fbar^{(3)} (m) \,, 
}} 
where we have rewritten $(-1)^m=1/ \cos (\pi
m)$. Assume that there are no relative cancellations of
residues, and that the contribution to the sum comes from a single
residue. The sum \fone\ arises from the poles at $m\in \Nop$.
In order to cancel the
poles of $\Gamma (m+1)$ for $m\in\Zop_-$, $\fbar^{(3)}(m)$ needs to
have a factor $1/\Gamma(m+1)$. Further, it can have poles at
$m\beta\in \Zop_-$ or $m\beta\in\Zop_+$, \ie, $\fbar^{(3)}_m \propto \Gamma(\pm\beta m)$. In fact, it has to have such a factor,
as otherwise, evaluating the contour integral for \ftwo\ would imply
that $\fbar^{(1)}_n$ vanishes, which is unphysical. 
Further, since $\beta<0$ the only consistent choice is
that the poles are at $m\beta\in \Zop_+$, as otherwise $\fbar^{(3)}_m$ would
be singular for the particular value $\beta=-1$. In summary we deduced that
\eqn\fthreeansatz{ 
\fbar^{(3)}(m) = \hat{f}^{(3)}(m)
{\Gamma(-m\beta)\over \Gamma(m+1)}\,, 
} 
with $\hat{f}^{(3)}(m)$ having no poles at $m\in\Zop$ or $m\beta\in \Zop$.

In evaluating the integral for \fthree
\eqn\ftwocontour{
\oint {dm\over 2\pi i} \, m {\Gamma (-m)\over \Gamma (1+ m\beta)} {\pi \over n^2 -
m^2 (1+\beta)^2} \fhat^{(3)}_m \,,
}
one obtains poles at $m=\pm n/(1+\beta)$. Assuming that the
contributions to the sums come only from one term, implies that $\fhat^{(3)}_m $ has to have
zeroes at all values $m$, such that $\eta m (1+\beta)\in\Zop_+$
for one of the signs $\eta=\pm 1$.
Thus, we may make the further ansatz
\eqn\fcheckansatz{
\fhat^{(3)}_m = {1\over \Gamma (-\eta m (1+\beta) +1)} \fcheck^{(3)}_m \,.
}
For the choice $\eta=1$\foot{With the choice $\eta=-1$, one would
encounter a pole at some positive real integer for some choice of
$\beta$. The non-vanishing contributions from this pole would lead to inconsistencies.}, the residue of \ftwocontour\ at $m= - n/(1+\beta)$
implies
\eqn\ftwosolved{
\fbar_n^{(2)} =  {\Gamma\left({n\over 1+\beta}\right) \over \Gamma \left(1- n
\left({\beta\over 1+\beta}\right)
\right) \Gamma (n+1)} \fcheck^{(3)}_{-{n\over 1+\beta}}
=  {\Gamma(-n {\alpha_3\over \alpha_2})\over \Gamma\left(-n\left(
{\alpha_3\over \alpha_2} +1\right) + 1\right)\Gamma(n+1)  }  \fcheck^{(3)}_{n{\alpha_3\over \alpha_2}}
 \,.}
The contour argument applies only if the integrand suitably falls off
at infinity (\cf\ \refs{\lss}). Invoking Stirling's formula,
the $m$-dependent part of the integrand in \ftwocontour\ has an
asymptotic behaviour given by
\eqn\ftwoasymp{\eqalign{
m^{-5/2} \beta^{m\beta } (1+\beta)^{m (1+\beta)} \fcheck^{(3)}(m)
&= m^{-5/2}
\alpha_1^{-m {\alpha_1\over \alpha_3}}
\alpha_2^{-m {\alpha_2\over \alpha_3}}
(-\alpha_3)^{-m}\fcheck^{(3)}(m)\cr
&= m^{-5/2} e^{-m\tau_0/\alpha_3} \fcheck^{(3)}(m)\,.
}}
Thus, $\fcheck^{(3)}$ must contain a factor $e^{m\tau_0/\alpha_3}$, but
could in principle also be proportional to $\sum_k a_k m^k$ for some
suitable powers $k$, which respect the required asymptotics. The same
asymptotic behaviour is obtained for the integrands for \fone\ and \fthree.
We shall fully determine $\fcheck$ below.

With the new ansatz \fcheckansatz, the residue for the sum in \fone\ becomes
\eqn\foneres{
\sum\limits_{m=1}^\infty {(-1)^m\over m} \sin (m \pi \beta )
\fbar^{(3)}_m = \pi \left.{d\fcheck^{(3)}(m) \over dm} \right|_{m=0} \,,
}
so that
\eqn\fcheckcondition{
\left.{d\fcheck^{(3)}(m) \over dm} \right|_{m=0}
 = {\tau_0\over \alpha_3} \,.
}
Now consider the integral for equation \ftwo,
\eqn\fonecontour{
(-1)^n \oint {dm\over 2\pi i}  {\pi\cot (\pi m)\over \cos (\pi m)} m  {\sin (m\pi
\beta)\over n^2-m^2\beta^2} {\Gamma (- m \beta) \over \Gamma(m+1)} \fhat^{(3)} (m)\,.
}
The only pole that has a non-trivial residue is located at $m=  n/\beta$, which results in
\eqn\fonesolved{
(-1)^{n+1}  {\pi \over 2 \beta^2} {\Gamma(- {n\over \beta})\over
\Gamma(n+1)} \fhat^{(3)}_{{n\over \beta}} =  {\pi \over 2\beta^2
} \fbar^{(1)}_n \,.
}
With \fcheckansatz, the function $\fbar^{(1)}$ is now determined as
\eqn\foneagain{
\fbar^{(1)}_n = (-1)^{n+1} {\Gamma(- {n\over \beta})\over \Gamma(-{n\over
\beta}
-n+1)\Gamma (n+1)} \fcheck^{(3)}_{n\over \beta}
=  {\Gamma (-{\alpha_2\over \alpha_1} n) \over \Gamma \left(-n
\left({\alpha_2\over \alpha_1}+1
\right)+1\right) \Gamma (n+1)} \fcheck^{(3)}_{n {\alpha_3\over \alpha_1}}
 \,,
}
where the reflection identity has been applied.
In order to further constrain the function $\fcheck$ we need to discuss
the last equation \ffour. Consider the term that gives rise to the
first term in \ffour, involving $\fbar^{(1)}_n$
\eqn\foneforall{
- \oint {dn\over 2\pi i} \, n  {
\fcheck^{(3)}({n\over\beta }) \Gamma (-n) \over n^2 -m^2\beta^2}
{\Gamma (n+ {n\over \beta}) \over \Gamma (1+ {n\over \beta})}\,.
}
The poles are determined as in appendix E of \refs{\GSstringint}. At
$n=0$ the pole is
\eqn\reszero{
\Res_{n=0} =
- {\fcheck^{(3)}(0) \over \beta (1+\beta) m^2}\,.}
At $n=m\beta$ the residue is
\eqn\resone{
\Res_{n=m\beta} =  - \fcheck^{(3)}(m)
{\Gamma(-\beta m)\Gamma((1+\beta)m) \over 2 \Gamma (1+m)} \,,
}
and for $n\in {\beta/(1+\beta)}\Zop_-$ the residues give rise to the sum
\eqn\restwo{
-\sum\limits_{k=1}^\infty \fcheck^{(3)}\left(-{ k\over 1+\beta}\right)
{1\over k^2 -m^2 (\beta+1)^2} {\Gamma({k\over \beta+1 })\over
\Gamma(k) \Gamma (1-{k\beta\over 1+\beta})}\,.
}
Comparison with \ffour\ yields the following additional condition on
$\fcheck^{(3)}$
\eqn\fcheckconds{\eqalign{
{\fcheck^{(3)}(0)\over \alpha_1 (1+\beta) m^2} &=
-{\alpha_3\over \alpha_1\alpha_2 m^2}\,,
}}
so that
\eqn\fcheckcon{
\fcheck^{(3)}(0)=1\,.
}
In order to fully determine $\fcheck$, recall that from the
asymptotical behaviour in \ftwoasymp\ we deduced that $\fcheck^{(3)} (m)=
\sum_k a_k m^k e^{m\tau_0/\alpha_3}$. Now, \fcheckcon\ and
\fcheckcondition\ imply that $a_0=1$ and $a_1 =0$. Since any higher
power of $k$ would alter the asymptotics such that the contour method
would not be applicable anymore, we conclude that
\eqn\fcheckfixed{
\fcheck^{(3)}(m) = e^{m\tau_0/\alpha_3} \,.
}
This is in agreement with \refs{\GSstringint}.
In summary, the solutions to the equations \fone-\ffour\ take the general form
\eqn\fsolved{
\fbar^{(r)}_m = \fcheck^{(r)}_{m}
{\Gamma \left(-m {\alpha_{r+1} \over \alpha_r}\right)   \over \Gamma (m+1) \Gamma
\left(-m \left({\alpha_{r+1} \over \alpha_r} +1\right) +1\right)} \,,
}
where we set
\eqn\fcheckr{
\fcheck^{(r)}_{m} = \fcheck^{(3)}_{m {\alpha_3\over \alpha_r}}
= e^{m\tau_0/\alpha_r} \,.
}
In particular, \fone-\ffour\ imply that the Neumann vectors are
given by
\eqn\Nmats{
-{1\over \alpha_r} \sqrt{m} \fbar^{(r)}_m =
\left(\left(A^{(r)}\right)^t \Gamma^{-1} B
\right)_m
= -  \Nbar_m^r \,,
}
which completes the contour method derivation for the flat-space cubic vertex. 


\newsec{The cubic vertex for the plane-wave}

Having illustrated the contour method, we are now ready to apply
it to derive the cubic vertex for the plane-wave string theory. Again,
we are interested in the bosonic Neumann coefficients. As in the case
of the open-closed vertex derived in \refs{\lss}, the cubic vertex
will turn out to be most concisely expressed in terms of
$\mu$-deformed Gamma-functions. In particular, the following functions will be useful
\eqn\morenewGamma{\eqalign{
\Gamma^{(r)}_\mu (z) &= e^{-\gamma \alpha_r \omega_z} \, {1\over \alpha_r z}\, \prod\limits_{n=1}^\infty
\left(
{n \over \omega_{rn} + \alpha_r \omega_z} \;
e^{\alpha_r\omega_z\over n}\right) \cr
&= \Gamma^I_{2 \mu \alpha_r} (\alpha_r z)\,,
}}
where $\omega_{rn}= \sqrt{n^2 + \alpha_r^2 \mu^2}$ and $\Gamma^I_\mu
(z)$ is the $\mu$-deformed Gamma-function defined in \refs{\lss}
(except $n$ as opposed to $\omega_n$ appears in the denominator of the
infinite product). We shall define the Gamma-function without a
superscript
\eqn\gamnoscripts{
\Gamma_{\mu}(z)= \Gamma^{(r)}_{\mu}(z) \,,\qquad \hbox{for } \alpha_r=1 \,.
}
A key property of these functions is that they satisfy a
generalization of the reflection identity of the Gamma-function
\eqn\refl{
\Gamma^{(r)}_{\mu}(z)  \Gamma^{(r)}_{\mu}(-z) = -{ \pi \over \alpha_r z \sin
(\pi \alpha_r z)} \; .}
Various properties of these functions, such as asymptotics in $z$ and in $\mu$ are
discussed in the appendices and in \refs{\lss}.


\subsec{Vertex equations}

The ansatz for the bosonic part of the plane-wave cubic vertex is as in \roughvertex.
The conditions on the Neumann matrices in the plane-wave case have
been derived in \refs{\spradlinvolovich}, which again
reduce to the problem of finding $f^{(3)}_m$ (denoted by $Y_m$ in \refs{\hssv}) such that
\eqn\pptosolve{
\sum_{n=1}^{\infty}(\Gamma_+)_{mn} \,  f^{(3)}_n =
\sum_{r=1}^3 \sum_{n=1}^{\infty} \left(A^{(r)} U^{(r)} \left(A^{(r)} \right)^t\right)_{mn} \, f^{(3)}_n
= B_m \,,
}
where
\eqn\Ur{
(U^{(r)})_{mn} = \delta_{mn} {(\omega_{rm}-\alpha_r \mu)\over m} \,.
}
The conventions are as in \refs{\spradlinvolovich,
\spradlinvolovichII, \hssv} and the relation to section 2 is by $f^{(r)}_n|_{\mu=0} = \fbar^{(r)}_n \sqrt{n}$ .

As in flat space, \pptosolve\ has the interpretation of continuity
conditions on the vertex at the interaction, \ie\ at $\tau=0$.
The strategy, which we shall pursue (and which is in contrast to
\refs{\hssv}) is to proceed as in flat space and stepwise
solve for $f^{(3)}_m$, \ie, to find solutions $f^{(r)}_m$ to the set of equations
\eqnn\stratagemone
\eqnn\stratagemtwo
$$
\eqalignno{
\sum_{p=1}^\infty \, f^{(3)}_p  A^{(r)}_{pm}  &= {\alpha_3\over \alpha_r}
\,  f^{(r)}_m &\stratagemone \cr
\sum_{p=1}^\infty \sum_{r=1}^3 {1\over \alpha_r} \, \left(A^{(r)} U^{(r)}\right)_{mp} f^{(r)}_p &=- B_m\,. &\stratagemtwo
}
$$
The Neumann matrices in the plane-wave case have been shown
\refs{\schwarz} to satisfy an analogous equation to \Nfactor,
\eqn\nmatdef{
\bar N^{rs}_{mn}=- {mn\alpha\over 1-4\mu\alpha K}{\bar
N^r_m\bar N^s _n\over \alpha_s\omega_{rm}+\alpha_r\omega_{sn}}
\,,}
where the Neumann vectors are
\eqn\nvecdef{\eqalign{ 
\bar{N}^r_m &=  \left(\,  (C^{-1}
C_r)^{1/2} U_r^{-1}\, f^{(r)} {1\over\alpha_r}\,
\right)_m  \cr 
&= \sqrt{\omega_{rm}\over m} \, {(\omega_{rm}+ \mu\alpha_r )\over m}
{1 \over\alpha_r} f^{(r)}_m \,,
}}
and
\eqn\ppK{
K=-{1\over 4} B^t \Gamma_+^{-1} B
\,.}
Further $C_{mn}=m\delta_{mn}$ ,
$(C_r)_{mn} = \omega_{rm} \delta_{mn}$ and the $f^{(r)}$'s are defined as above.

Our strategy is now to apply the contour method to the sums in
\stratagemone\ and \stratagemtwo.
From equation \stratagemone\ one can again deduce the pole structure for
$f^{(3)}(m)$. Assuming that the residues of the equations in
\stratagemone\ come from a single pole the conditions are that
$f^{(3)}(m)$ has zeroes for
$m\in -\Nop$ as well as $m(1+\beta)\in \Nop_0$ and has poles for
$m\beta\in \Nop$. Thus, this fixes the pole and zero structure of the
solution, however not the explicit functional dependence. The latter
is determined by \stratagemtwo. For this, note the $\omega_{rp}$ term entering
$U^{(r)}$. As discussed in \refs{\lss}, the integrals along the branch cuts
that are present due to the square root will not contribute to the contour integral
corresponding to the sum in question, if the $\mu$ and $p$ dependences
are all packaged together into $\omega_{rp}$ and the integrand is odd
in the imaginary part. Thus, in view of
\stratagemone, one is lead to the following explicit realization of
the poles and zeroes in $f^{(3)}$
\eqn\fthreeansatz{
f^{(3)}(m) = \tilde{f}^{(3)}_m
{\Gamma_{-\mu \beta}(-\beta m)\over  \Gamma_{\mu (1+\beta)}(-(1+\beta)m) \,
\Gamma_{\mu}(m) }\,,
}
where the particular choice of deformation parameters for the Gamma
functions is chosen, in order to ensure that all branch cuts
coincide. The function $\tilde{f}^{(3)}_m$ is determined 
much in the same way as we explained in detail for the flat-space discussion.
Furthermore, one has to ensure that the all-$\mu$
solution reproduces the right flat-space limit. Note also, that the
contour method requires that the integrand falls off suitably at
infinity, so that the RHS of \contone\ vanishes. In the flat-space this
discussion relied on applying the Stirling formula to the Gamma
functions. For the plane-wave case, it will be relevant that the
$\mu$-deformed Gamma-functions satisfy an analogous Stirling formula,
which is proven in appendix B.


\subsec{Solutions to the vertex equations}

From the discussion in the last subsection, which in particular lead to the form \fthreeansatz, one obtains the following ansatz for the function $f^{(3)}(m)$, which has
the correct flat-space limit. It further satisfies
${\partial f^{(3)}\over \partial \mu}\big{|}_{\mu=0}=\tau_0 f^{(3)}
$, which follows from equation (30) of \refs{\hssv}\foot{Note this
motivates the factor $1/ \omega_m$ as opposed to $1/ (\omega_m+ \mu)$
which both have the correct flat space limit.}.
The ansatz reads
 \eqn\ppfthree{ f^{(3)}_m = {m^2 \over 2
\omega_m} B_m y(1-y) e^{\tau_0(\mu-\omega_m)} {
\Gamma^{(1)}_{\mu}(m)\Gamma^{(2)}_{\mu}(m) \over \Gamma_{\mu}(m)}
M(0^+) \,, }
where we fixed as in \refs{\hssv} the gauge
\eqn\gaugechoice{
 \alpha_1=y\,,\quad \alpha_2=1-y\,,\quad
\alpha_{3}=-1 \,, 
} and 
\eqn\Bdef{ 
B_{m}= {2\over \pi } {\sin
(m\pi (1-y)) \over y (1-y) m^{3/2}} \,, } as well as \eqn\Mdef{
M(z)= {\Gamma_{\mu} (z) z \over \Gamma_{\mu y} (yz) yz \,
\Gamma_{\mu (1-y)} ((1-y)z) (1-y)z} \,. 
}
The factor $M(0^+)$ is computed from \fthreeansatz\ by imposing the
equation \stratagemtwo\ and thus is crucial in order to reproduce the correct residues.
Note also that $M(0^+)\to 1$ as $\mu\to 0$. So, more explicitly we
have 
\eqnn\ppfthreeagaina 
\eqnn\ppfthreeagainb
$$
\eqalignno{
f^{(3)}(m) &= {\sqrt{m}\over \pi} \sin (m \pi (1-y)) {e^{\tau_0 (\mu -\omega_m)} \over \omega_m } \,
{\Gamma_{\mu y}(ym) \Gamma_{\mu (1-y)}((1-y)m)  \over
\Gamma_{\mu}(m) } \, M(0^+) &\ppfthreeagaina \cr
&=- {e^{\tau_0 (\mu -\omega_m)} \over \omega_m } \, {1\over \sqrt{m}
(1-y)}\,  {\Gamma_{\mu y}(ym)\over  \Gamma_{\mu (1-y)}(-(1-y)m) \,
\Gamma_{\mu}(m) } \, M(0^+) \,. &\ppfthreeagainb
}
$$
Evaluation of the contour integrals corresponding to \stratagemone\ for $r=1$ and $r=2$, which have only
non-trivial residues at $m=-n/y$ and $m=-n/(1-y)$, respectively,
results in
\eqnn\ppfone
\eqnn\ppftwo
$$
\eqalignno{
f^{(1)}(n) &= -{ (-1)^{n} e^{\tau_0 (\mu + \omega_{n\over y})}\over \omega_{n\over y} \sqrt{n} (1-y)}
{\Gamma_{\mu} ({n\over y }) \over \Gamma^{(2)}_{\mu}({n\over
y})\,  \Gamma^{(1)}_{\mu} ({n \over y})} \, M(0^+) &\ppfone \cr
f^{(2)}(n) &=  { e^{\tau_0 (\mu + \omega_{n\over 1-y})}\over
\omega_{n\over 1-y} \sqrt{n} y} \,
{\Gamma_{\mu} ({n\over 1- y }) \over \Gamma^{(1)}_{\mu }({ n\over
1-y}) \, \Gamma^{(2)}_{\mu} ({n \over 1-y})} \, M(0^+)  \,. &\ppftwo
}
$$
Note that in evaluating the contour integrals for each value for $r$
the branch cuts coincide, the integrand being odd
along the cuts, which is of course crucial 
for the applicability of the contour method. 
These solutions can be put into the closed form 
\eqn\closedformsolution{
f^{(r)}_m= { e^{\tau_0 (\mu+\omega_{{m\over \alpha_r}})}
\over \sqrt{m} (-\alpha_{r} -\alpha_{r+1}) \, 
\omega_{m\over \alpha_r}} 
{\Gamma_{\mu}^{(r+1)}\left(- {m\over \alpha_r}\right) \over 
\Gamma_{\mu}^{(r)}\left({m\over \alpha_r}\right)
\Gamma_{\mu}^{(r-1)} \left({m\over \alpha_r}\right)} 
\, M(0^+)\,,
}
which beautifully generalizes the corresponding expression in
flat-space \fsolved\ taking into account the difference in conventions.

Next \stratagemtwo\ needs to be checked. Thus consider the following contour
integral, which corresponds to the sum of the $r=1$ term in
\stratagemtwo\ (\cf\ flat space analysis) 
\eqn\pponeU{
- M(0^+) \oint {dn \over 2\pi i} \, {2y (-1)^{m+1} \sqrt{m}
\sin(m\pi y)\over n \sin(\pi n)} {e^{\tau_0 (\mu + \omega_{n\over
y})}\over n^2 - m^2 y^2} {(\omega_{n\over y} -\mu)\over
\omega_{n\over y}} {\Gamma_{\mu (1-y)}(-{n\over y}(1-y)) \over
\Gamma_{\mu}(-{n\over y}) \Gamma_{\mu y} (n)}\,. 
} 
The residues at $n \in\Nop$ give the $r=1$ term, the residues at $n(1-y)/y
\in\Nop$ give the $r=2$ term and the residue at $n= -my$ gives the
$r=3$ term in \stratagemtwo. We are left only with the integral
around the branch cut, which runs from $n=iy\mu$ to $n=-iy\mu$.
Since the integrand is odd along either side of the cut the line
integrals vanish -- however the integrand is actually singular at
$n=0$ and thus we are left with two semicircular contours on
either side of the branch cut, which we will call $C_+$ and $C_-$, see
figure 1.
One might be tempted to think this contribution is just the
residue of the integrand  at $n=0$ however things are not quite as
simple as this, since the integrand is not defined at this point
with our choice of branch cut. What we can do though is take the
residue of the factors not depending on $\omega_n$ and then on the
right side (\ie, along $C_+$) take the limit $n \to 0^+$ of the
rest and on the left take the limit $n \to 0^-$.

\fig{The contours $C_\pm$.}{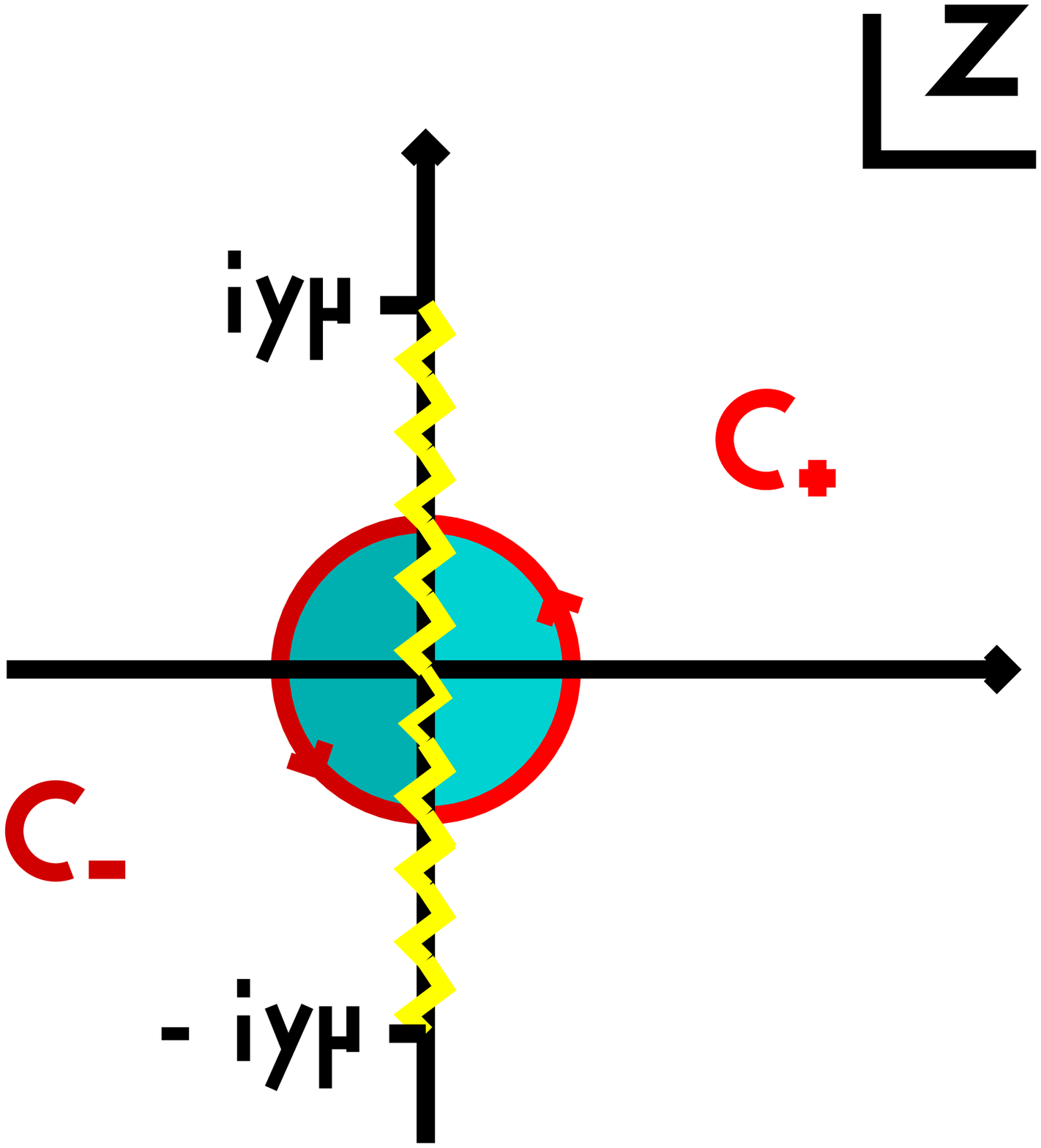}{2.1truein}

In the present case, the only contribution comes
from $C_-$ (note the integrand is not regular at the branch points, however the integral around them still vanishes, in fact goes
as $O(\epsilon^{1/2})$ where $\epsilon$ is the radius of the contour
around a branch point, see \refs{\lss}). We find that
the contribution from $C_-$ is
\eqn\cplusint{
- {2(-1)^m \sin m\pi y \over \pi m^{3/2} y(1-y)} M(0^+)M(0^-) \; .}
It is easy to verify that $M(0^+)M(0^-)=1$, which leaves the
contribution from $C_-$ to be exactly $B_m$ thus completing the proof.

Finally, we need to compute the quantity $K$, which is defined by
\eqn\Kdef{
K= -{1\over 4} B^t \, \Gamma_+^{-1}\, B =-{1\over 4} \sum_{n=1}^\infty B_n f^{(3)}_n \,. 
}
For the computation, we need to consider the contour integral
\eqn\Kcalc{
-{ 2M(0^+) \over y(1-y)} \oint {dm \over 2\pi i} {
 M(-m)e^{\tau_0(-\omega_m +\mu)} \over m\omega_m} \, .}
The residues at $-m \in\Nop$ give $-4K$ and since there are no other
 poles we are left with the integrals around the branch cut. There is
 a singularity in the integrand at $m=0$ which on both sides looks
 like a simple pole. Computing we get
\eqn\Kcplus{
-{ 2M(0^+) \over y(1-y)} \int_{C_+} {dm \over 2\pi i} {
 M(-m)e^{\tau_0(-\omega_m +\mu)} \over m\omega_m} = -{M(0^+)M(0^-)
 \over \mu y(1-y)}\, ,}
and
\eqn\Kcminus{
-{ 2M(0^+) \over y(1-y)} \int_{C-} {dm \over 2\pi i} {
 M(-m)e^{\tau_0(-\omega_m +\mu)} \over m\omega_m} = {M(0^+)^2
 e^{2\tau_0 \mu} \over \mu y(1-y)} \, .}
Therefore we find that
\eqn\Kexpr{
K= {1 \over 4y(1-y) \mu} \left( M(0^+)^2 e^{2\mu\tau_0} - 1\right) \; ,}
which we see has the correct flat space limit, $K(\mu=0)= \tau_0/(2y (1-y))$.

We should emphasize that in this application of the contour method of \refs{\lss} it has been crucial that
the circular integrals around the branch points, as well as the
integrals at infinity vanish -- the latter was shown by applying
the generalization of Stirling's formula of appendix B.

\subsec{Comments}

A couple of remarks are in order, in particular in view of the
comparison with the solution obtained in \refs{\hssv}. First
recall that the solution $Y_m(\mu, y)$ in
\refs{\hssv} is to be compared with our solution $f^{(3)}_m$.
In this comparison, one will observe that the
solution in eq. (52) of \refs{\hssv} only agrees with \ppfthree,
if $\alpha_3\to -\alpha_3$ as well as the sign of the $r=3$ term
in the exponent is flipped, \ie, if 
\eqn\ymcorrect{ 
Y_m
(\mu, y) = \exp \left[(\mu-\omega_m)\tau_0 + \sum_{r=1}^2 (\phi_r
- \phi_{mr}) - (\phi_3 - \phi_{m3})\right] {m\over 2\omega_m} B_m
\,, } 
with $\alpha_3$ replaced by $1$. Note that the solution as
it stands in \refs{\hssv} would be divergent for integral $m$, and
thus seems inconsistent\foot{The authors of \refs{\hssv} have
confirmed the sign discrepancies in the equation in question. We would like
to thank Y. He and M. Spradlin for discussions on this point.}.


\newsec{Large-$\mu$ asymptotics}

The
large-$\mu$ asymptotics of expressions in  plane-wave string
theory are of the foremost interest in the context of the BMN correspondence, as these are to be compared with perturbative (in $\lambda'$)
gauge theory data. Given our expressions for the Neumann matrices, the
only non-trivial input in determining these asymptotic expressions are
the large-$\mu$ asymptotics of the $\mu$-deformed
Gamma-functions, which we derive in appendix C. In applying these one has to keep in
mind that the asymptotic formulae only apply to $\Gamma^{(r)}(z)$ for
$|\hbox{arg}(z)|<\pi$, thus before applying the asymptotics to the
functions $f^{(r)}(m)$ one has to ensure that upon suitable
application of the reflection identities this condition is
satisfied for the arguments.

First note that $M(0^+)$ has the asymptotics, using the
$\Gamma$-function asymptotics of the appendix C, given by \foot{As
in \refs{\hssv}, $A(\mu) \approx B(\mu)$ means $A(\mu)= B(\mu)
+O(e^{-|\alpha_r| \mu})$.}
\eqn\Mnaughtasymp{ 
M(0^+) \approx
e^{-\tau_0 \mu} \sqrt{1\over 4 \pi \mu y (1-y) } \,. }
Applying
these asymptotics upon \ppfone\ and \ppftwo, we obtain
\eqn\frlargemu{\eqalign{ 
f^{(1)}_n &\approx {(-1)^n\over 2\pi
\sqrt{2\mu } y (1-y)} {\sqrt{n}\over \omega_{n\over y}
\sqrt{\omega_{n\over y} + \mu}} \cr 
f^{(2)}_n &\approx {1\over 2 \pi \sqrt{2\mu} y (1-y)} {\sqrt{n}\over \omega_{n\over (1-y)}
\sqrt{\omega_{n\over (1-y)} + \mu}} \,.} }
For $f^{(3)}(n)$, the
asymptotic formula is only applicable to \ppfthreeagaina,
resulting in
\eqn\frlargemuotherway{ 
f^{(3)}_m \approx {1\over
\pi} \sin( m\pi \alpha_{2}) {\alpha_3\over \sqrt{m}} {1\over \sqrt{2\mu}
\alpha} { \sqrt{\, \omega_{-{m\over \alpha_3}}+\mu\, }\over
\omega_{m\over \alpha_3} } \,. }
This expansion agrees precisely
with \refs{\hssv}. The Neumann vectors
\eqn\neuvecs{\eqalign{  
\bar{N}^r_m &= \sqrt{\omega_{rm}\over m} \, {(\omega_{rm}+
\mu\alpha_r )\over m} f^{(r)}_m  {1\over\alpha_r}\cr 
&= \sqrt{\omega_{rm}\over m} \, {m\over
(\omega_{rm}- \mu\alpha_r )} f^{(r)}_m {1\over\alpha_r}
 \,, }}
thus have the asymptotic behaviour for $r=1,2$
\eqn\neuasym{
\bar{N}^r_m \approx -{(-1)^{r(n+1)} \alpha_r^{1/2}  \over
2\pi \alpha \sqrt{2\mu} m} {\sqrt{(\omega_{rm} + \mu \alpha_r)}
\over \sqrt{\omega_{rm}}} 
\,, } 
as well as for $r=3$
\eqn\neuasymthree{ 
\bar{N}^3_m \approx -{1\over \pi} \sin (m\pi
\alpha_2) {1\over \sqrt{2\mu} \alpha} \sqrt{ |\alpha_3|\over
\omega_{3m} (\omega_{3m}-\mu\alpha_3)}
\,. }
Finally we may study
the large-$\mu$ asymptotics of $K$. Inspecting \Kexpr\ we see that
we only need \Mnaughtasymp, which leads to
\eqn\Klargemu{ 
K \approx -{ 1 \over 4y(1-y) \mu} + { 1 \over 16\pi y^2(1-y)^2
\mu^2} \; .}
These agree with the findings of \refs{\hssv}.


\subsec{Exponential corrections}

In this section we will concentrate on explicitly extracting the
first exponential correction to the Neumann vectors we have
derived, thus going further than the results given in
\refs{\hssv}. These corrections could have interesting
implications for the gauge theory.

The problem reduces to finding the large-$\mu$ exponential
corrections to the deformed Gamma-functions. Here we give a brief
argument and one can find a more rigorous derivation in appendix
D. In appendix D we show the following key result
\eqn\loggammaexpfinal{ 
{ \partial \over \partial \mu} \log
\Gamma^{(r)}_{\mu}(z)_{\rm exp}= - { \sqrt{|\alpha_r| \mu} \over
\omega_z } e^{-2 \pi |\alpha_r|\mu} \left[ 1+ O\left( {1 \over \mu}
\right) \right] 
\,.} 
If one does an integration with respect to $\mu$ by
parts (\ie, integrate the $e^{-2 \pi |\alpha_r| \mu}$), then we can
prove that
\eqn\exploggamma{ 
\log \Gamma_{\mu}^{(r)}(z)_{\rm
exp} = { \sqrt{|\alpha_r| \mu} \over 2\pi
|\alpha_r| \omega_z} e^{-2\pi |\alpha_r| \mu}   
\left[ 1+ O \left( {1\over \mu} \right) \right]
\,,} 
the details of which are provided in appendix D. This essentially
means that the remaining integral simply contributes
to the $O(1/\mu)$ part. 
Now it is a simple matter of some algebra to deduce the
corrections to the Neumann vectors. We find the following result
\eqn\threegamma{\eqalign{ 
{ \Gamma_{\mu}^{(1)}(z)
\Gamma_{\mu}^{(2)}(z) \over \Gamma_{\mu}(z) } &= { \sqrt{2 \pi}
e^{\omega_z \tau_0} \over \sqrt{y(1-y)}} { \sqrt{\omega_z+ \mu}
\over z} \bigg\{ 1+ { \sqrt{\mu} \over 2\pi \omega_z } \bigg(
{e^{-2 \pi y \mu}\over\sqrt{y}} \left[ 1+ O\left( {1 \over \mu} \right) \right]
\cr &+ {e^{-2 \pi (1-y) \mu}\over \sqrt{1-y}} \left[ 1+ O\left( {1 \over \mu}
\right) \right] - e^{-2 \pi \mu} \left[ 1+ O\left( {1 \over \mu}
\right) \right] \bigg) \bigg\} 
\,.}}
Using this it is
straightforward to deduce the corrections to $f^{(r)}_m$.

Curiously, it can be shown that the exponential corrections are
related to the Casimir energy of a two-dimensional massive boson
on a cylindrical world-sheet with periodic boundary condition. The
Casimir energy is given by \refs{\bgg,\taka}
\eqn\casimir{
\Delta_{\mu\alpha_r}=-{1\over
(2\pi)^2}\sum_{p=1}^\infty\int_0^\infty dt\,
e^{-p^2t-\pi^2\alpha_r^2\mu^2/t}
\,,} 
using which it can be easily
shown that the exponential corrections for the deformed Gamma-function satisfies the following relation 
\eqn\logdercas{
{\partial
\log \Gamma^{(r)}_{\mu}(z)_{\rm exp}\over
\partial \mu}={1\over \omega_z \alpha_r}{\partial
\Delta_{\mu\alpha_r}\over \partial \mu}={1\over \omega_z \mu}{\partial
\Delta_{\mu\alpha_r}\over \partial \alpha_r}
\,.}
$\alpha$ parametrizes the length of the
world-sheet. Since the $\alpha$-derivative of the energy can be interpreted as
pressure, these corrections probably arise due to the tidal
forces in this background \refs{\bgg}.
The exact physical significance of this result, particularly on the
gauge theory side needs to be explored.

It is tempting to extend the analysis applied for the first term in
\exploggamma\ in order to extract the full series of exponential
corrections. In this paper, we shall content ourselves by giving a
conjectural expression for the series \exploggamma,
\eqn\expconjecture{
\log \Gamma_{\mu}^{(r)}(z)_{\rm
exp}= 
\sum_{n=1}^{\infty} e^{-2\pi n |\alpha_r|\mu}\, {\sqrt{|\alpha_r|\mu}\over 2\pi
|\alpha_r|\sqrt{n}\omega_z} \left( 1+O\left({1\over \mu} \right)\right)
\,,}
where the conjecture is, that the $O(1/ \mu)$ terms do not lead to any mixing of the
exponentials in the series. 
We provide some arguments in favour of this conjecture in appendix D.1.

Using the above conjecture and upon expanding the exponential, in general the exponential
corrections can be written schematically as
\eqn\genexp{
\sum_{n=1}^\infty g_n(\mu,\alpha_r)e^{-2\pi n
\mu|\alpha_r|}\,,} 
which suggests the interpretation of these
quantities in terms of a series of non-perturbative corrections.
Note, that in deriving the
exponential corrections we made use of the asymptotic expansion of
the Bessel function $K_{\nu}(x)$. It is well known that this expansion is not
Borel summable. The exact significance of this for the dual gauge
theory needs to be understood.


\newsec{Plane-wave Neumann matrix manual}

The purpose of this section is to summarise our main results, in order
to facilitate the comparison with gauge theory calculations for which the
Neumann
matrices are essential.
We give the expressions for the
Neumann matrices valid for all $\mu$, the large-$\mu$ expansion as
well as the first exponential corrections, which we have determined
explicitly. The subsequent
exponential terms are only conjectural and can be obtained from
\expconjecture.

So, to summarize, we have expressed the plane-wave Neumann matrices as
\eqn\Nmatdef{
\bar N^{rs}_{mn}=- {mn\alpha\over 1-4\mu\alpha K}{\bar
N^r_m\bar N^s _n\over \alpha_s\omega_{rm}+\alpha_r\omega_{sn}}
\,,}
where the Neumann vectors $\bar N^r_m$ are related to the functions
$f^{(r)}_m$ via
\eqn\Ntof{
\bar{N}^r_m 
=  \sqrt{\omega_{rm}\over m} \, {(\omega_{rm}+ \mu\alpha_r )\over m}
{1\over\alpha_r} f^{(r)}_m 
\,.}
We have determined the explicit form of these functions in
\closedformsolution, and furthermore $K$ was
computed in \Kexpr. Putting all this together, we obtain the Neumann
matrices for all $\mu$-values to be
\eqn\Neumannfinal{\eqalign{
\bar N^{rs}_{mn} \, =\  & - \, 
{4 \over m n} \ 
e^{{\tau_0 \left(\omega_{m\over \alpha_r} +
\omega_{n\over \alpha_s}\right)}}\times \cr
& \times\ 
{\alpha \over (\alpha_r +\alpha_{r+1}) \, (\alpha_s + \alpha_{s+1}) }\ 
{ (\omega_{rm} + \mu
\alpha_r)(\omega_{sn}+\mu \alpha_s) \over
(\omega_{rm}\omega_{sn})^{1/2}\, 
(\alpha_s\omega_{rm}+\alpha_r\omega_{sn})}\ \times \cr 
& \times\  \left(
{\Gamma_{\mu}^{(r+1)}\left(- {m\over \alpha_r}\right) \over 
\Gamma_{\mu}^{(r)}\left({m\over \alpha_r}\right)
\Gamma_{\mu}^{(r-1)} \left({m\over \alpha_r}\right)} 
\right)\ 
\left(
{\Gamma_{\mu}^{(s+1)}\left(- {n\over \alpha_s}\right) \over 
\Gamma_{\mu}^{(s)}\left({n\over \alpha_s}\right)
\Gamma_{\mu}^{(s-1)} \left({n\over \alpha_s}\right)} 
\right)\,, 
}}
where the $\mu$-deformed Gamma-functions are defined in \morenewGamma\
and \gamnoscripts.
Applying the large-$\mu$ asymptotics for the deformed Gamma-functions 
obtained in appendix C one can extract the large-$\mu$ behavior, which
is of interest for comparison with the gauge theory. 
We have given the explicit forms of the large-$\mu$ Neumann vectors in
section 4, and the asymptotics for \Nmatdef\ are straight-forwardly
obtained from \neuasym,
\neuasymthree\ and \Klargemu. 

The first term in the exponential corrections to these large-$\mu$ asymptotics, which had
so far not been determined, follow from the exponential corrections of the
Gamma-functions in \threegamma, together with \Nmatdef.


\newsec{Conclusions}

In this paper we have derived the bosonic Neumann matrices in plane-wave string theory
using the contour method developed in \refs{\lss}, which allows to
express the result in terms of $\mu$-deformed Gamma-functions.
This approach not only simplifies the derivation of the Neumann
matrices and their large-$\mu$-asymptotics, the latter being in agreement
with \refs{\hssv}, but allows to extract exponential corrections,
\ie, terms of $O(e^{-2\pi |\alpha_r|\mu})$. 
We have derived an explicit form for the first term in these
exponential corrections, and provided a conjectural formula for the
leading order in $1/\mu$ terms in the full exponential series. A few
open questions and remarks are in order.

\smallskip \noindent $\bullet$ The dual gauge-theoretical
interpretation of the exponential corrections has certainly so far
been elusive. The explicit form for the first
term in the exponential corrections, which we derived, provides some
explicit quantity that could be compared to the gauge theory. In terms of the
effective 't Hooft coupling, $\lambda'=1/(\mu p^+\alpha')^2$, the
result is proportional to $e^{-2\pi/\sqrt{\lambda'}}$. In particular,
these should correspond to non-perturbative effects, which are remotely reminiscent of
contributions that arise from D-branes in string theory. A vital question that hereby arises is then: What objects on the gauge theory side could be
attributed such corrections?

\smallskip \noindent $\bullet$ The exponential corrections were
shown to be related to the Casimir energy of a massive
two-dimensional boson on a cylinder. Is this merely a mathematical
coincidence or can it be attributed more physical significance?

\smallskip \noindent $\bullet $ It would be very interesting to
use the results obtained in the present paper as well as in \refs{\lss} in
order to compute scattering of closed and open strings. In particular,
the open string cubic vertex is of course closely related to the closed cubic
vertex and could be used in order to compute scattering of open
strings in orientifold theories or with D-branes, such as the ones
constructed in \refs{\bgg, \gg, \ggss,\asnvs}. Furthermore, these 
should be compared to gauge
theoretical computations including operators dual to D-branes, such as
in \refs{\giantgraviton, \berensteinnarain, \berenstein}.

\smallskip \noindent $\bullet $ As a mathematical curiosity, it is
conceivable that the contour method could be used more generally to
derive integral transforms, in the same way that the present paper
gives a systematic way to obtain the integral transform used in
\refs{\hssv} to derive the cubic vertex. 

\smallskip \noindent $\bullet $ Finally, it is tempting to conjecture that all flat space amplitudes, which can be
expressed in terms of Gamma-functions, can be carried over in the
plane-wave background by replacing them with suitable $\mu$-deformed Gamma-functions.


\vskip 1cm \centerline{{\bf Acknowledgments}}

\noindent We thank Michael B. Green, Sean Hartnoll, Yang-Hui He, Minxin Huang, Stefano Kovacs,
S. Prem Kumar and Mark Spradlin for very useful discussions. JL is
supported by EPSRC. SSN thanks the University of Pennsylvania and the IAS, Princeton, for
hospitality during the final stages of this work. AS
acknowledges financial support from the Gates Cambridge Trust and
the Matthews' scholarship of Gonville and Caius College,
Cambridge. 
\pano


\appendix{A}{Notations and Conventions}

The following definitions have been used in the main body of the paper
\eqn\AB{\eqalign{
A^{(1)}_{mn} &= { 2 \over \pi} (-1)^{m+n+1} \sqrt{mn} { \beta \sin(m \pi
\beta) \over n^2 - m^2 \beta^2} \cr 
A^{(2)}_{mn} &= { 2 \over \pi} (-1)^{m+1} \sqrt{mn} { (\beta +1) \sin(m \pi
\beta) \over n^2 - m^2 (\beta+1)^2} \cr 
A^{(3)}_{mn} &= \delta_{mn} \cr 
B_m &= { 2 \over \pi}{ \alpha_3 \over \alpha_1 \alpha_2} (-1)^{m+1} {
\sin( m \pi \beta) \over m^{3/2}} \,,
}}
which arise in the Fourier mode expansion of the vertex
equation. Further it is useful to define
\eqn\CU{\eqalign{
C_{mn} &= m\, \delta_{mn} \cr
(C_r)_{mn} &= \omega_{rm}\, \delta_{mn} \cr
(U^{(r)})_{mn} &= \delta_{mn} {(\omega_{rm}-\alpha_r\mu)\over m}
\,.}}
Here, $\beta = \alpha_1 / \alpha_3$. We will mostly work in with the
choice $\alpha_1 =y$ and $\alpha_2 = 1-y$ and hence $\alpha_3 =-1$, as
in \refs{\hssv}.


\appendix{B}{Generalisation of Stirling's formula}

In this section we analyse the large $z$ asymptotics of the Gamma
function $\Gamma_{\mu}(z)$. Recall this is defined as \refs{\lss}
\eqn\Gammamu{
\Gamma_{\mu}(z) = {e^{-\gamma \omega_z} \over z} \prod_{n=1}^{\infty}
{n \over \omega_z + \omega_n} e^{\omega_z / n} \; .}
Using the Weierstrass definition of $\Gamma(z)$ implies
\eqn\logGammas{
\log \left( {\Gamma_{\mu}(z) \over \Gamma(z)} \right) =-
\gamma(\omega_z - z) + \sum_{n=1}^{\infty} \log \left( { z+n \over
\omega_z +\omega_n} \right) + {\omega_z - z \over n} \; .}
Note that $\lim_{z \to \infty} (\omega_z - z) =0$. This allows us to
deduce that
\eqn\Gammamuasylog{
\lim_{z \to \infty} \log \left( {\Gamma_{\mu}(z) \over \Gamma(z)}
\right) = 0 \,,}
and therefore that
\eqn\Gammamuasy{
\Gamma_{\mu}(z) \sim \Gamma(z) \sim \sqrt{2\pi} z^{z-1/2}e^{-z} 
\,,}
as $z \to \infty$ and of course for $|\arg z| < \pi$.


\appendix{C}{Large-$\mu$ asymptotics of the deformed Gamma-functions}

The large-$\mu$ asymptotics are derived in a similar fashion as in the
appendices of \refs{\lss}, applying various techniques of \refs{\WW}.
Taking the log of both sides of \morenewGamma\ and differentiating
with respect to $\mu$ leads to
\eqn\diffGamma{{\partial \over \partial \mu}\log
\Gamma_\mu^{(r)}(z)= {\alpha_r\mu\over
\omega_z}\left[\sum_{n=1}^{\infty} \left({1\over
n}-{1\over \omega_{rn}} \right) -\gamma \right]\,.}
So we need to consider the asymptotics of
\eqn\defS{S= \sum_{n=1}^{\infty}\left({1\over
n}-{1\over \omega_{rn}}\right)\,.}
Differentiating both sides with respect to $\mu$ implies
\eqn\diffS{{
\partial S\over \partial \mu}=\sum_{n=1}^\infty{\alpha_r^2
\mu \over \omega_{rn}^3}\,.}
Using the results in the appendix of \refs{\lss}
\eqn\eqStwo{
\sum_{n=1}^\infty{1\over \omega_{rn}^3} =-{1\over 2(\alpha_r
\mu)^3}+{1\over (\alpha_r \mu)^2}+O(e^{-|\alpha_r|\mu})
\,,}
which after integrating leads to
\eqn\eqSthree{
S={1\over 2\mu\alpha_r}+\log \mu+c(\alpha_r)+O(e^{-|\alpha_r|\mu})
\,,}
where $c(\alpha_r)$ is a constant of integration. Differentiating
with respect to $\alpha_r$ leads to
\eqn\diffalpha{
{\partial S\over \partial \alpha_r}=-{1\over 2\mu\alpha_r^2}+{\partial
c\over \partial \alpha_r}+O(e^{-|\alpha_r|\mu})
\,.}
One should now differentiate $S$ with respect to $\alpha_r$ and then take the
large-$\mu$ limit of the resulting expression to compare with this
one. This leads to
\eqn\eqSfour{{\partial S\over \partial
\alpha_r}=\sum_{n=1}^\infty\left({\mu^2\alpha_r\over
\omega_{rn}^3}\right)=-{1\over 2\alpha_r^2\mu}+{1\over \alpha_r}+O(e^{-|\alpha_r|\mu})\,.}
Now comparing with equation \diffalpha\ we get
\eqn\calpha{{\partial c\over \partial \alpha_r}={1\over \alpha_r}\,,}
implying that
\eqn\csol{c(\alpha_r)=\log \alpha_r +c\,,}
and therefore
\eqn\Ssol{
S={1\over 2\mu\alpha_r}+\log(\mu\alpha_r) +c +O(e^{-|\alpha_r|\mu})
\,.}
Substituting this into \diffGamma\ and then integrating with respect to $\mu$ we
arrive at
\eqn\loggammamuasy{\eqalign{
\log \Gamma_{\mu}^{(r)}(z) &= \alpha_r \omega_z \left( c-\gamma
-1 +\log(\mu\alpha_r) \right) + {1 \over 2} \log(
\omega_z +\mu) \cr &+  z\alpha_r \log \left( { \omega_z + z \over \mu}
\right) + K(z,\alpha_r) +O(e^{-|\alpha_r|\mu}) 
\,,}}
where $K(z,\alpha_r)$ comes from integrating with respect to $\mu$. This
function can be determined by taking the large-$\mu$ asymptotics of ${
\partial \over \partial z} \log \Gamma^{(r)}_{\mu}(z)$ and comparing to the
$z$ and $\alpha_r$ derivatives of \loggammamuasy, which we will do next.
Taking the partial derivative with respect to $z$ of \loggammamuasy\ leads
to
\eqn\diffznewGammaasy{{\partial\over\partial z}\log\Gamma_\mu^{(r)}(z)
\approx {z \over
2(\mu+\omega_z)\omega_z}+(c-\gamma-1+\log \alpha_r\mu){\alpha_r z\over
\omega_z}+\alpha_r\log \left( {z+\omega_z\over \mu} \right)+{z\alpha_r\over
\omega_z}+{\partial K\over \partial z}\,,}
while the $z$ logarithmic derivative of \morenewGamma\ leads to
\eqn\diffznewGammaexact{
-\gamma\alpha_r {z\over \omega_z}-{1\over
z}-{\alpha_r z\over \omega_z}\sum_{n=1}^{\infty}\left({1\over
\omega_{rn}+\alpha_r\omega_z} -{1\over n}\right)
\,.}
Taking the limit $\mu \to \infty$ in both equations results in the condition
\eqn\diffKz{
{\partial K\over \partial z}=-{1\over z}
\,.}
Now taking the partial derivative with respect to $\alpha_r$ of \loggammamuasy\
one obtains
\eqn\diffanewGammaasy{
{\partial\over\partial
\alpha_r}\log\Gamma_\mu^{(r)}(z) \approx (c-\gamma
+\log (\alpha_r\mu )) \omega_z+z\log \left({z+\omega_z\over \mu}\right)
+{\partial K\over
\partial \alpha_r}\,,}
while the $\alpha_r$ logarithmic derivative of \morenewGamma\
implies
\eqn\diffanewGammaexact{-\gamma\omega_z-{1\over\alpha_r}-\sum_{n=1}^\infty\left({1\over\omega_{rn}+\alpha_r\omega_z}
\left({\alpha_r\mu^2\over\omega_{rn}}+\omega_z\right)-{\omega_z\over
n}\right)\,.} 
Comparing the above two equations at $z=0$ and using
\Ssol\ leads to the condition 
\eqn\diffKa{
{\partial K\over
\partial \alpha_r}=-{1\over 2 \alpha_r}\,.} Therefore we conclude
that \eqn\solK{ K(z,\alpha_r)= -\log z - {1 \over 2} \log \alpha_r
+ c' \,.} 
Finally we have the desired asymptotic expression for
the deformed Gamma-functions for $|\arg z| < \pi$
\eqn\logGammamuasyfinal{
\eqalign{ \log \Gamma_{\mu}^{(r)}(z) &=
\alpha_r \omega_z \left( c-\gamma -1 +\log(\mu\alpha_r) \right) +
{1 \over 2} \log( \alpha_r\omega_z +\alpha_r\mu) \cr &+  z\alpha_r
\log \left( { \omega_z + z \over \mu} \right) - \log(z\alpha_r)
+c' +O(e^{-|\alpha_r|\mu}) 
\,.}} 
The constants $c$ and $c'$ can
actually be determined, and we will in fact need $c'$ explicitly. To determine these constants we employ the large
$z$-asymptotics formula derived in appendix B which is valid for
all $\mu$. For large $z$, the RHS of \logGammamuasyfinal\
(ignoring the $O(e^{-|\alpha_r|\mu})$ contribution) is asymptotic
to \eqn\largezasy{ \left( \alpha_r z -{1 \over 2} \right) \log
(\alpha_r z) + \alpha_r z( c-\gamma -1 +\log 2) + c' \; .}
Comparing this to Stirling's formula (which is valid for all
$\mu$) we see that we must have $c= \gamma -\log 2$ and $c' = \log
\sqrt{2 \pi}$.


\appendix{D}{$O(e^{-\mu})$ corrections}

We shall now derive the exponential corrections in the large-$\mu$
expansion of the deformed Gamma-functions. 
In equation \diffS\ , the $O(e^{-\mu})$ term is given by
\eqn\diffSe{
{\partial S_e\over \partial \mu}=2\alpha_r^2\mu\int_0^\infty {ds\over
\mu^2\alpha_r^2}
e^{-s}\sum_{n=1}^{\infty}e^{-{n^2\pi^2\mu^2\alpha_r^2\over s}}
\,,}
which can be written in terms of the modified Bessel function of the
second kind as
\eqn\diffSeK{
4\pi|\alpha_r|\sum_{n=1}^\infty n K_1(2n\pi\mu|\alpha_r|)
\,,}
where we have used the integral represenation $K_{1}(x) = {1 \over x}
\int_0^{\infty} dt \, e^{-t-x^2/4t}$.
Integrating with respect to $\mu$ leads to
\eqn\Sesol{
S_e=-2\sum_{n=1}^\infty K_0(2|\alpha_r|\mu n\pi)
\,.}
Thus the $O(e^{-\mu})$ terms in \diffGamma\ are given by
\eqn\diffGammae{
F(z,\mu) \equiv -2\sum_{n=1}^\infty {\alpha_r\mu\over \omega_z}K_0(2|\alpha_r|\mu n \pi)\,.}
The large-$\mu$ behaviour of this quantity is readily deduced from
that of the Kelvin functions\foot{Note $K_{\nu}(x) =  \sqrt{ {\pi \over
2x}} e^{-x} (1+ O(1/x))$.}, which gives
\eqn\Flargemu{
F(z,\mu) = - { \sqrt{|\alpha_r| \mu} \over \omega_z } e^{-2 \pi |\alpha_r|
\mu} \left[ 1+ O\left( {1 \over \mu} \right) \right] 
\,.}
Thus we are left with evaluating $\int F(z,\mu)\,d\mu$.
We now prove the following formula crucial for the
evaluation of the above integral
\eqn\intexp{
I(z)=\int_{\mu}^{\infty} dt { \sqrt{t} \over \sqrt{t^2 + z^2} } e^{-\xi
t} = {\sqrt{\mu} e^{-\xi \mu} \over \xi \omega_z } \left[ 1+ O \left(
{1 \over \mu} \right) \right] \; .}
We will prove the formula for real $z$ as this is all we will
need. The argument is elementary and goes as follows. Integrate by
parts to give
\eqn\parts{
I(z)= {\sqrt{\mu} e^{-\xi \mu} \over \xi \omega_z } - {1 \over 2 \xi}
\int_{\mu}^{\infty} dt e^{-\xi t} { (z^2 -t^2) \over \sqrt{t}
(z^2+t^2)^{3/2}} \; .}
Since $|z^2 - t^2| < z^2 + t^2$ for real $z$, we have
\eqn\ineq{
\left| { (z^2 -t^2) \over \sqrt{t} (z^2+t^2)^{3/2} } \right| < { 1 \over \sqrt{t}
  (z^2+t^2)^{1/2} } \leq { 1 \over \sqrt{\mu}
  (z^2+\mu^2)^{1/2} } \,,}
wherefore
\eqn\intapprox{
\left| {1 \over 2 \xi}
\int_{\mu}^{\infty} dt e^{-\xi t} { (z^2 -t^2) \over \sqrt{t}
(z^2+t^2)^{3/2}} \right| <  { e^{-\xi \mu} \over 2\xi^2 \sqrt{\mu}
\omega_z } \,.}
Hence we have proven \intexp. Note we have not restricted $\mu$ in
this proof at all. Thus one might expect to extend this to complex $z$
for large-$\mu$. Using this we conclude that
\eqn\Fintapprox{
\int_\mu^\infty F(z,\mu)\,d\mu={\sqrt{|\alpha_r| \mu}
\over 2\pi |\alpha_r |\omega_z}e^{-2\pi |\alpha_r|\mu} \left[ 1+O\left( {1\over
\mu} \right) \right]\,. }

Finally we comment on the connection to \refs{\hssv}. For this, we make use of the following integral representation
\eqn\intrepK{
K_0(x\mu)= \int_0^{\infty} dt \, { e^{-x \sqrt{t^2+\mu^2}} \over
\sqrt{t^2 +\mu^2} } \,,}
valid for $x>0$, see \refs{\GR}, to express the whole of the exponential corrections
in a different form. Using this we may sum $S_e$, given by
\Sesol, which implies
\eqn\sesum{
S_e = -2 \int_0^{\infty} dt \, {1 \over \sqrt{t^2+\mu^2} [ e^{2\pi
\alpha_r \sqrt{t^2+\mu^2}}-1]} \,.}
If we change variables to $\mu s = \sqrt{t^2+\mu^2}$ we obtain
\eqn\sefinal{
S_e = - 2  \int_1^{\infty} ds \, {1 \over \sqrt{s^2 -1}}
{1 \over e^{2\pi \mu\alpha_r s} -1} \,,}
which is a closed 
expression for the $O(e^{-\mu})$ terms of the sum \defS. The
exponential corrections to the $\mu$-deformed Gamma-functions are
readily obtained from \diffGamma. 
Incidentally an equivalent formula to \sefinal\ appears in \refs{\hssv}.


\subsec{Conjecture for subsequent exponential corrections}

In analogy to the derivation of the first term in the exponential corrections
one should be able to compute the full series \genexp. There are
various subtleties in determining this, in particular related to the
approximation of the $O(1/\mu)$ term in \parts. We shall now present
some arguments which allow us to conjecture the exact expression for
the leading order terms, \ie, of $O(1/\mu)$. 
So, we wish to compute the expansion of the following term into a series
of exponential corrections $O(e^{-2\pi n |\alpha_r|\mu })$
\eqn\wannado{ 
\int_{\mu}^{\infty} d\mu\,  {\mu\over \omega_z}\sum_{n=1}^\infty K_0(2|\alpha_r|\mu n \pi)\,.
}
First, recall the asymptotic series for the Kelvin function
(\refs{\watsonbessel}, VII., 7.34)
\eqn\kelvins{
K_0 (x) = \sqrt{\pi \over 2x} e^{-x} \left(  \sum_{m=0}^{p-1} {c_m \over
(2x)^m} + (-1)^p R_p  \right) \,,
}
where the error term $R_p$ for large $p$, such that $x= p/2+ \sigma$
with $|\sigma|<1 $, is given by
\eqn\kelverror{
R_p \sim 2\sqrt{x\over \pi} {e^{-2x}\over p} \left(\half + O\left({1\over p}\right)\right)\,.
}
Applying this to the Kelvin function appearing in \wannado, implies that
\eqn\moreexp{
\log \Gamma_{\mu}^{(r)}(z)_{\rm
exp} = \sum_{n=1}^{\infty}
e^{-2\pi n |\alpha_r|\mu}
{\sqrt{|\alpha_r|\mu}\over 2\pi
|\alpha_r|\sqrt{n}\omega_z} \left( 1
+O\left({1\over \mu}\right) \right)\,,
}
where it may be of use to point out that the $R_p$-terms do not
contribute to the leading term in $1/\mu$, and thus can be
disregarded.
If it now can be ensured that $O(1/\mu)$ does not contain terms like
$e^{-\mu}$ then we can retain the first term,
as there is no mixing at the same order of the coefficient. However, to
make this statement precise, a better approximation of the $O(1/\mu)$
terms in \parts\ would have to be derived.


\listrefs
\bye